\documentclass{elsart5p}

\usepackage{graphicx}
\usepackage{epsfig}

\usepackage{lineno}

\newcommand{\degr}{^{\circ}}

\begin{document}

\begin{frontmatter}
\title{HADES - Hydrophone for Acoustic Detection at South Pole}

\author{Benjamin Semburg}
\ead{semburg@physik.uni-wuppertal.de}
\author{for the IceCube collaboration}\footnotemark
\address{Bergische Universit\"at Wuppertal, Fachbereich C ---
  Mathematik und Naturwissenschaften, 42097 Wuppertal, Germany}

\begin{abstract}
The South Pole Acoustic Test Setup (SPATS) is located in the upper part of the optical neutrino observatory IceCube, currently under construction. SPATS consists of four strings at depths between 80\,m and 500\,m below the surface of the ice with seven stages per string. Each stage is equipped with an acoustic sensor and a transmitter. Three strings (string A-C) were deployed
  in the austral summer 2006/07. SPATS was extended by a fourth string (string D) with
  second generation sensors and transmitters in 2007/08. One second generation
  sensor type HADES (Hydrophone for Acoustic Detection at South Pole) consists
  of a ring-shaped piezo-electric element coated with polyurethane.
  The development of the sensor, optimization of acoustic transmission by acoustic
  impedance matching and first in-situ results will be discussed.
\end{abstract}

\begin{keyword}
acoustic neutrino detection, SPATS, HADES
\PACS 43.58.Bh 
\sep 43.58.$+$z 
\sep 93.30.Ca 
\end{keyword}
\end{frontmatter}

\footnotetext{http://www.icecube.wisc.edu}

\section{Motivation}
\label{sec:motivation}

The optical neutrino observatory IceCube is currently under construction at the geographic South Pole, Antarctica. IceCube strings with 60 optical modules between 1450\,m and 2450\,m (17\,m spacing) per string are installed vertical in holes, drilled with hot water \cite{Ach 06}.\\

\begin{figure}[htbp]
\begin{center}
\begin{minipage}[t]{5,3 cm}
\includegraphics[width=1.0\textwidth]{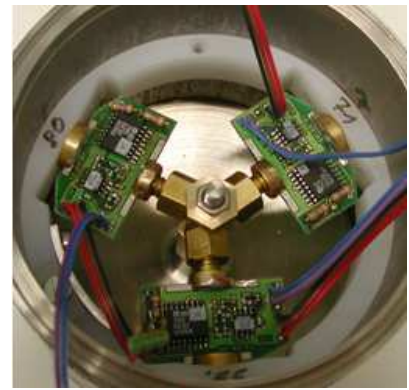}
\caption{Inside view of one SPATS sensor housing used at SPATS strings A-C \cite{Bse 06}.}
\label{fig:spats stage}
\end{minipage}
\end{center}
\end{figure}

In the austral summer season 2006/2007 three strings of the South Pole Acoustic Test Setup (SPATS) were installed after the IceCube strings in the upper part of IceCube holes down to 400\,m depth.
Each of these three SPATS strings consists of 7 stages
and each stage contains one transmitter steel housing with a transmitter and one sensor steel housing. Three piezo ceramic elements with amplifiers are located inside the steel housing and pressed to the inner wall with a preload screw (see Figure \ref{fig:spats stage}) \cite{Bse 06}.\\

SPATS was deployed to answer some major questions on the way to a hybrid (acoustic, radio and optical) neutrino detector: attenuation length of acoustic waves, sound speed profile in the deep polar ice, absolute noise level and the origin of transient noise events.\\

To improve the first results (published in \cite{Bse 07}) and to reduce the systematic uncertainties, a fourth SPATS string (string D) was deployed in the austral summer season 2007/2008.
String D also consists of 7 stages with one transmitter and one sensor housing per stage. The spacing is different in comparison to strings A-C (string D is instrumented between 140\,m and 500\,m depth) and the transmitters and sensors are improved \cite{Van 08}. All sensors on string D, except at 190\,m and 430\,m depth, contain three piezo elements, which are fixed inside the steel housing with the assistance of a ring instead of a preload screw to achieve better coupling to the inner wall.\\

HADES A and HADES B are located at 190\,m and 430\,m depth on string D, respectively. The motivation for HADES (\underline{H}ydrophone for \underline{A}coustic \underline{De}tection at \underline{S}outh Pole) was to develop a sensor which is an alternative to the standard SPATS sensor design with a different dynamic range, different systematic effects and optimized signal transmission. It is adapted to connect to a modified SPATS sensor housing for easy integration into the existing setup. For the HADES sensors we used only one ring-shaped piezo element per sensor with a different amplifier board in comparison to the SPATS sensors. The piezo element is located outside of the steel housing and we used resin as coating material.

\section{HADES}
\label{sec:hades}

\begin{figure}[htbp]
\begin{center}
\begin{minipage}[t]{8,5 cm}
\includegraphics[width=1.0\textwidth]{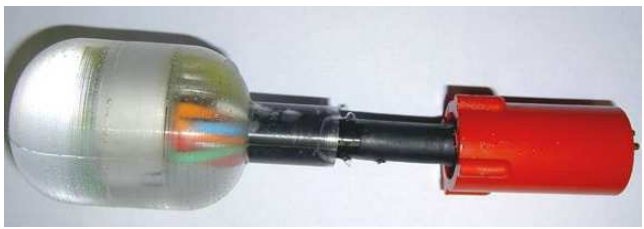}
\caption{HADES sensor with piezo-ceramic element and two-step amplification in transparent resin (left) and the 8 pin connector (right).}
\label{fig:hades sensor}
\end{minipage}
\end{center}
\end{figure}

The HADES sensor consists of a piezo-electric element (ring-shaped piezo\footnote{http://www.ferroperm-piezo.com} with inner diameter: 20\,mm, outer diameter: 24\,mm and height: 15\,mm) that is connected to an amplifier board (amplifier: Ti TL072) with two step amplification. The amplifier sends a differential signal via an 8-pin male connector (MCIL8M\footnote{http://subconn.com}). For protection against water and ice the piezo and the amplifier are coated with resin (see Figure \ref{fig:hades sensor}).\\

Both HADES sensors are connected to SPATS via  modi-fied SPATS sensor housings. This sensor housing contains only a voltage converter board and an 8-pin female connector (MCBH8F\addtocounter{footnote}{-1}\footnotemark) at the bottom of the sensor housing to plug in the HADES sensor.\\

\subsection*{Acoustic impedance matching}

Acoustic signals inside the South Pole ice have to cross several media until they can be detected with acoustic sensors: For example from the bulk ice into the hole ice through the sensor housing to the sensor itself. During these propagation processes the acoustic signal gets refracted and partially reflected at media interfaces.

In the following only the transition effect from ice to the sensor housing will be discussed.\\  

To maximize the received signal, the transmission coefficient ($T$) from the South Pole ice to the sensor has to be maximized:

\begin{equation}\label{eq:transmission}
  T = \frac{4 \cdot Z_{\rm{ice}} \cdot Z_{\rm{sensor}} \cdot \cos(\alpha_1) \cdot \cos(\alpha_2)}{[Z_{\rm{ice}} \cdot \cos(\alpha_2) + Z_{\rm{sensor}} \cdot \cos(\alpha_1)]^2}
\end{equation}

The acoustic impedance $Z$ of a medium is given by the product of the density and the speed of sound inside the medium:

\begin{equation}\label{eq:impedanz}
  Z = \rho \cdot v_{\rm{sonic}}
\end{equation}

\begin{figure}[htbp]
\begin{center}
\begin{minipage}[t]{6,8 cm}
\includegraphics[width=1.0\textwidth]{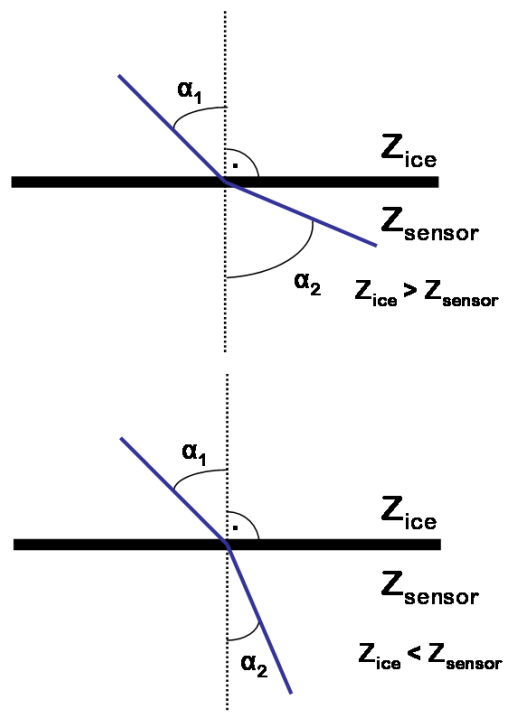}
\caption{Refraction with respect to acoustic impedance properties.}
\label{fig:refraction}
\end{minipage}
\end{center}
\end{figure}

The acoustic impedance corresponds to the refraction index in optics and Snellius law (see Figure \ref{fig:refraction}):

\begin{equation}\label{eq:snellius}
  Z_{\rm{ice}} \cdot \sin (\alpha_1) = Z_{\rm{sensor}} \cdot \sin (\alpha_2)
\end{equation}

Four different resins (epoxy, polyester, hard and soft polyurethane) were tested for low temperature qualification (down to $-\,85\,\degr$C) and for acoustic impedance properties relative to the acoustic impedance of South Pole ice.\\

Test rods of each resin passed the $-\,85\,\degr$C air temperature freezer test. These test rods were slowly cooled down to $-\,85\,\degr$C air temperature and back to room temperature. After two cycles no deformation or damage were visible.\\

On the other hand, the epoxy and polyester failed at low temperatures in the consecutive test as casting material for the piezo element. The test was repeated with four piezo elements coated with these four different resins. All resins didn't have the possibility to tighten homogeneously when they cover a piezo element. The inner stresses inside epoxy and polyester at $-\,85\,\degr$C caused the effect that these two materials bursts and that they were separated from the piezo element.\\

\begin{figure}[htbp]
\begin{center}
\begin{minipage}[t]{8,5 cm}
\includegraphics[width=1.0\textwidth]{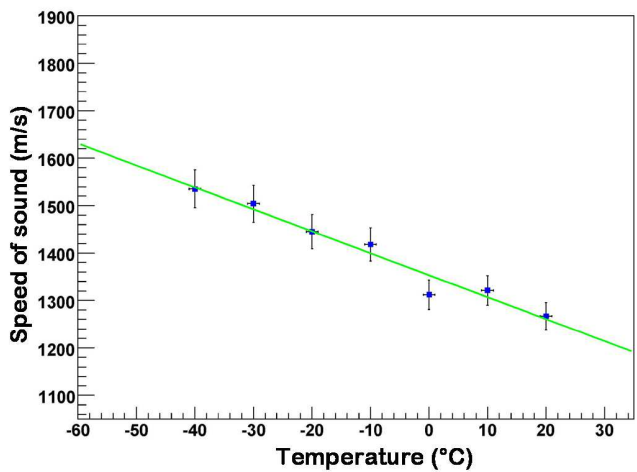}
\caption{Speed of sound vs. temperature for hard polyurethane.}
\label{fig:speed of sound}
\end{minipage}
\end{center}
\end{figure}

The speed of sound for the acoustic impedance (equation \ref{eq:impedanz}) was determined through the length of the test rod and the signal transit time. The signal transit time was measured as a function of the temperature in a climate chamber\footnote{climate chamber MK 53, http://binder-world.com}. Piezo ceramic elements infused in hard polyurethane served as transmitter and receiver at the ends of the rod. The sent signal was a Gaussian shaped pulse with a center frequency of 50\,kHz. See Figure \ref{fig:speed of sound} for the speed of sound vs. temperature of hard polyurethane.\\

The lowest achievable temperature of the climate chamber is $-\,40\,\degr$C, whereas the temperature of the South Pole ice is $-\,50\,\degr$C. Therefore the speed of sound needs to be linearly extrapolated from the data (see Figure \ref{fig:speed of sound})\\

The acoustic impedance of South Pole ice at $-\,50\,\degr$C is \cite{Pri 05}:

\begin{equation}\label{eq:Eis -50}
  Z_{\rm{acoustic}}^{\rm{ice}}(-\,50\,\degr \rm{C}) = 3606\,\frac{\mathrm{kNs}}{\mathrm{m}^3}
\end{equation}

The soft polyurethane has a lower density compared to hard polyurethane, which influences the acoustic impedance and finally the HADES sensors are coated by a two component hard polyurethane\footnote{Resin (Biothan 1770s) and catalyst (Biodur 330) from http://www.modulor.de/shop} (hard PU) resin. \\

With the given density the acoustic impedance of the hard PU can be calculated with equation \ref{eq:impedanz}:

\begin{equation}\label{eq:hartesPu -50}
  Z^{\rm{hard PU}}(-\,50\,\degr \rm{C}) =
  (1836\,\pm\,88)\frac{\mathrm{kNs}}{\mathrm{m}^3}
\end{equation}

With the assumption of normal incidence ($\alpha_1 = \alpha_2 = 0\degr$) equation \ref{eq:transmission} becomes:

\begin{equation}\label{eq:transmission_simply}
  T = \frac{4 \cdot Z_{\rm{ice}} \cdot Z_{\rm{sensor}}}{[Z_{\rm{ice}} + Z_{\rm{sensor}}]^2}
\end{equation}

With the calculated impedances of equation \ref{eq:hartesPu -50}, \ref{eq:Eis -50} and the simplified equation of normal signal transmission (equation \ref{eq:transmission_simply}) the signal transmission from South Pole ice to hard PU at $-\,50\,\degr$C can be calculated to T$^{\rm{hard PU}}_{\rm{ice}}$($-\,50\,\degr$C) $\approx$ 89\,\%.\\

All other possible transmission effects due to different media, mentioned above, and the complex geometry of the sensors are currently under theoretical study and laboratory experiments are under preparation.

\section{Commissioning}
\label{sec:commissioning}
\begin{figure}[htbp]
\begin{center}
\begin{minipage}[t]{8,0 cm}
\includegraphics[width=1.0\textwidth]{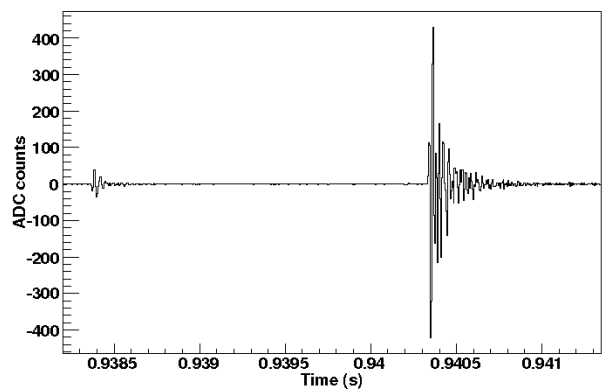}
\caption{Intra-stage event detected with HADES B in 430\,m depth.}
\label{fig:intra-stage event}
\end{minipage}
\end{center}
\end{figure}

Intra-stage runs were taken, starting 24\,h after deployment. At these runs a transmitter of a certain stage is fired and the sensor of the same stage received the signal. Figure \ref{fig:intra-stage event} shows an intra-stage event of HADES B on string D at 430\,m depth. The (acoustic) pre-pulse, caused by the charging process when the transmitter electronics generate the pulse, and the acoustic pulse, caused by the transmitter itself, are clearly visible.\\

\begin{figure}[htbp]
\begin{center}
\begin{minipage}[t]{8,5 cm}
\includegraphics[width=1.0\textwidth]{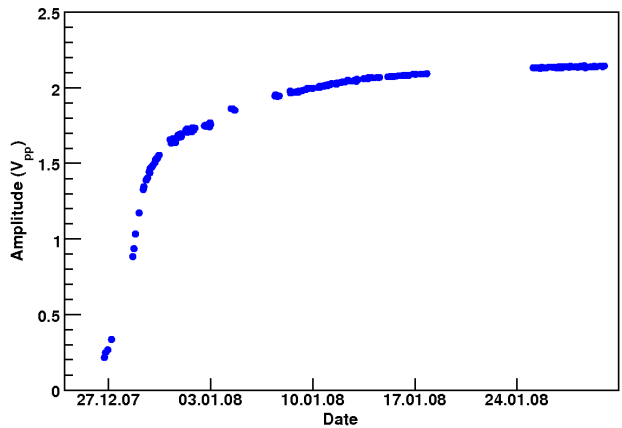}
\caption{Measured peak to peak amplitudes vs. time from intra-stage runs with HADES B.}
\label{fig:freezeIn}
\end{minipage}
\end{center}
\end{figure}

Figure \ref{fig:freezeIn} shows the measured maximum voltage (peak to peak) of each intra-stage run versus time for HADES B during freeze in of the string. Possible reasons for the increase in received signal strength could include:\\

\begin{itemize}
\item Increase in sensitivity due to temperature decrease since the piezo element is more sensitive at low temperatures.\\

\item Better acoustic coupling between sensor \& ice or/and transmitter \& ice.\\

\item Geometry changes of the hole during freezing.\\
\end{itemize}

The amplitude increase by a factor of $\approx$\,10 is the first in-situ measurement at this depth, pressure and temperature. As mentioned above the increase in amplitude could be caused by the desired sensitivity increase due to acoustic impedance matching.

\section{Triggered transient events}
\label{sec:triggered events}

After the conference the SPATS DAQ was slightly modified and currently all four SPATS strings are taking transient noise data with three different sensor channels per string.

\begin{figure}[htbp]
\begin{center}
\begin{minipage}[t]{8,0 cm}
\includegraphics[width=1.0\textwidth]{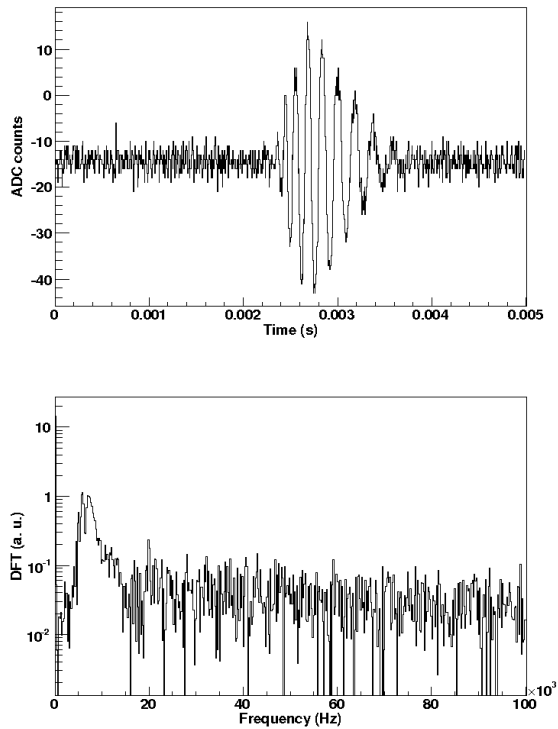}
\caption{Triggered transient event (upper panel) with Discrete Fourier Transform (lower panel) detected by HADES A in 190\,m depth.}
\label{fig:transient event}
\end{minipage}
\end{center}
\end{figure}

Figure \ref{fig:transient event} shows a typical transient event in HADES A data. The visible signal between 5\,kHz and 15\,kHz in the frequency domain of the shown transient event is visible in almost all analyzed events detected by HADES A. The signal is also visible in all found coincidence events within string D transient data (one HADES channel and two different SPATS channels).

Analysis for coincidences with all strings and vertex reconstruction are in progress.

\section{Summary}
\label{sec:conclusion and outlook}

A fourth SPATS string (string D) was successfully deployed in 2007/2008. String D contains revised SPATS sensors and stronger transmitters. With HADES A \& B a new type of sensor was deployed. The HADES sensor consists of a piezo element coated by resin and for that reason we have different coupling effects between sensor and ice in comparison to SPATS sensors.\\

We presented the first acoustic in-situ measurement of a possible freezing process. It seems that the sensitivity of the HADES sensor increases by a factor of $\approx$\,10 due to change in environment conditions (temperature and pressure). Laboratory measurements to verify these effects are under preparation.

\section{Acknowledgments}
\label{sec:acknowledgments}

The deployment and success of the South Pole Acoustic Test Setup would not have been possible without the friendly hospitality at the National Science Foundation Amundsen-Scott South Pole Station, Antarctica.\\

We thank the ANTARES working group at University Erlangen-N\"urnberg for their support in acoustic sensor development.\\

This work was supported by German Ministry for Education and Research.


\end{document}